\documentclass[12pt,preprint]{emulateapj}

\shorttitle{An extremely luminous $z>7$ galaxy in ECDFS}
\shortauthors{Hsieh et al.}

\begin{document}

\title{The Taiwan ECDFS Near-Infrared Survey:
Very Bright End of the Luminosity Function at $z>7$} 

\author{Bau-Ching Hsieh\altaffilmark{1}, Wei-Hao Wang\altaffilmark{1},
Haojing Yan\altaffilmark{2}, Lihwai Lin\altaffilmark{1},
Hiroshi Karoji\altaffilmark{3,4}, Jeremy Lim\altaffilmark{1,5},
Paul Ho\altaffilmark{1,6}, Chao-Wei Tsai\altaffilmark{7}
}

\altaffiltext{1}{Institute of Astrophysics \& Astronomy, Academia Sinica,
P.O. Box 23-141, Taipei 106, Taiwan, R.O.C.}

\altaffiltext{2}{Department of Physics and Astronomy,
University of Missouri, Columbia, MO 65211, USA}

\altaffiltext{3}{National Astronomical Observatory of Japan, 2-21-1 Osawa, 
Mitaka, Tokyo, 181-8588, Japan}

\altaffiltext{4}{Institute for the Physics and Mathematics of the Universe, 
University of Tokyo, 5-1-5 Kashiwanoha, Kashiwa,277-8583 Japan}

\altaffiltext{5}{Department of Physics, 
University of Hong Kong, Pokfulam Road, Hong Kong}

\altaffiltext{6}{Harvard-Smithsonian Center for Astrophysics, 
60 Garden Street, Cambridge, MA 02138, USA}

\altaffiltext{7}{Infrared Processing and Analysis Center,
California Institute of Technology, 770 South Wilson Avenue,
Pasadena, CA 91125, USA}

\begin{abstract}
The primary goal of the Taiwan ECDFS Near-Infrared Survey (TENIS) is
to find well screened galaxy candidates at $z>7$ ($z'$ dropout) 
in the Extended Chandra Deep Field-South (ECDFS). 
To this end, TENIS provides relatively deep $J$ 
and $K_s$ data ($\sim25.3$ ABmag, $5\sigma$) 
for an area of $0.5\times0.5$ degree.
Leveraged with existing data at mid-infrared to optical wavelengths,
this allows us to screen for the most luminous high-$z$ objects,
which are rare and thus require a survey over a large field to be found.
We introduce new color selection criteria
to select a $z>7$ sample with minimal contaminations 
from low-$z$ galaxies and Galactic cool stars;
to reduce confusion in the relatively low angular resolution IRAC images,
we introduce a novel deconvolution method to measure the IRAC fluxes
of individual sources.
Illustrating perhaps the effectiveness at which
we screen out interlopers, we find only one $z>7$ candidate, TENIS-ZD1.
The candidate has a weighted $z_{phot}$ of 7.8,
and its colors and luminosity indicate 
a young (45M years old) starburst galaxy with
a stellar mass of $3.2\times10^{10}$ M$_\odot$. 
The result matches with
the observational luminosity function analysis
and the semi-analytic simulation result
based on the Millennium Simulations, 
which may over predict the volume density for high-$z$ massive galaxies.
The existence of TENIS-ZD1, 
if confirmed spectroscopically to be at $z>7$,
therefore poses a challenge to current theoretical models for
how so much mass can accumulate in a galaxy at such a high redshift.
\end{abstract}

\keywords{galaxies: formation --- galaxies: evolution ---
galaxies: high-redshift} 

\section{Introduction}\label{introduction}
Finding objects at high redshifts is of fundamental importance
in observational cosmology.
By doing so, we are probing closer to 
the first galaxies and black holes in the universe,
and therefore ever closer to the initial conditions 
for the formation of galaxies and structures.
At the present time, we do not know when galaxies first formed, 
nor how quickly they can grow in mass during their extreme youth.
Because $z\sim7$ is the epoch that marks the end of the reionization,
finding objects at $z>7$ will provide crucial constraints
on the source of ionizing photons, as well as in
the early assembly history of galaxies and black holes.

The last decade or so has bore witness to an explosive growth in the discovery
of galaxies or galaxy candidates at $z\sim 6 - 10$, 
pushing our understanding of galaxy formation
to ever earlier epochs after the Big Bang.
Many efforts have been made using various techniques
\citep[e.g.,][]{taniguchi2005,kashikawa2006,iye2006,shimasaku2006,stark2007,
mannucci2007,ota2008,ota2010,richard2008,
bouwens2008,bouwens2009a,bouwens2010a,bouwens2010b,bouwens2010c,oesch2009,
oesch2010a,oesch2010b,castellano2010,mclure2009,mclure2010,mclure2011,
bunker2010,hickey2010,henry2007,henry2008,henry2009,bradley2008,zheng2009,
capak2011,wilkins2010,wilkins2011,sobral2009,gonzalez2010,ouchi2009,ouchi2010,
yan2011a},
but all can be roughly separated into three categories:
(1) ground-based narrow-band imaging searches for Ly$\alpha$ emitters (LAEs)
(2) space-based (HST/WFC3) ultra-deep narrow-field surveys
for Lyman Break galaxies (LBGs)
(3) ground-based deep and wide-field surveys for LBGs.
The continuum emissions of the LAEs provided by category (1) 
are usually too faint to be detected 
even photometrically;
and so their UV star-formation rate and the continuum luminosity function (LF)
of these objects cannot be derived.
After HST/WFC3 was launched, 
the sample size of category (2) increased rapidly and
now dominates the $z>6$ sample.
Given the limited survey field sizes, however,
these studies mainly focus on very faint samples 
as the probability of finding luminous but rare objects are small.
By contrast, surveys corresponding to category (3) are capable 
of finding rare bright sources 
that provide the strongest leverage in constraining the bright end of the LF.

Most of the current $z>7$ sample are provided by
surveys in category (2) and (3). 
The vast majority of these $z>7$ candidates have been identified based on
their photometric colors and do not have spectroscopic confirmations.
Contamination by intervening objects may be severe
in the currently known sample.
There are two major sources of contamination,
heavily reddened low-$z$ galaxies and Galactic cool stars.
For the samples provided by category (2), 
low-$z$ galaxies can be well resolved by HST/WFC3
and so their identification is straightforward.
A Galactic cool star, however, may have both similar colors and morphology
as a compact galaxy at $z>7$,
and so contamination from cool stars remains an issue.
Samples provided by category (3) suffer from these major contaminations.
This issue has been discussed in detail in many previous studies
\citep[e.g.,][]{ouchi2009,capak2011}, 
and has been recognized to be difficult to resolve.
In this paper, we use Spitzer IRAC photometry
to introduce additional criteria for
significantly reducing both the abovementioned contaminations 
in the category (3) studies for selecting $z>7$ candidates.

As we utilize the Spitzer IRAC photometry in our analysis,
we require a field with deep multi-wavelength data 
including Spitzer IRAC observations.
For this purpose, the Extended Chandra Deep Field South 
\citep[ECDFS,][]{lehmer2005} is ideal. 
The ECDFS is a $0.5\times0.5$ degree field that surrounds both
the Chandra Deep Field South (CDFS)
and the Great Observatories Origins Deep Survey South (GOODS-S) field.
The ECDFS has been studied at many wavelengths 
in addition to the original Chandra X-ray data. 
In the UV, the Deep Imaging Survey (DIS) for
the Galaxy Evolution Explorer ultraviolet satellite 
\citep[GALEX,][]{martin2005} observed the entire ECDFS.
In the optical, the Galaxy Evolution from Morphology and SEDs 
\citep[GEMS,][]{rix2004,caldwell2008} project
used the Advanced Camera for Surveys (ACS) on the Hubble Space Telescope (HST) 
to cover nearly the entire ECDFS. 
The ECDFS has also been obserbed in many ground-based optical and near-infrared 
imaging projects, e.g.,
the Classifying Objects through Medium-Band Observations,
a spectrophotometric 17-filter survey 
\citep[COMBO-17,][ optical only]{wolf2001} program, 
Multiwavelength Survey by Yale-Chile 
\citep[MUSYC,][]{gawiser2006,taylor2009,cardamone2010}
including 10 O/NIR broad-bands and 18 deep topical medium-bands,
and the NIR survey using ISAAC on VLT in the GOODS-S field.
In the infrared and far-infrared, 
most of the relevant data have been obtained using Spitzer. 
The Spitzer IRAC/MUSYC Public Legacy in the ECDFS survey 
\citep[SIMPLE,][]{damen2011}
provides deep IRAC data at wavelengths of 3.6, 4.5, 5.8, and 8.0$\mu{m}$
with a $10'\times15'$ ultra-deep GOODS-S IRAC data in the center
\citet{dickinson2003},
and the Far-Infrared Deep Extragalactic Legacy survey 
\citep[FIDEL,][]{dickinson2007}
took MIPS images at wavelengths of 24, 70, and 160$\mu{m}$
to cover nearly the entire ECDFS.
At radio wavelengths, 
the ECDFS has been covered by the VLA 1.4GHz Survey \citep[][]{kellerman2008}.

Like in previous surveys in category (3), 
we start by using the Lyman-break technique to find LBGs at $z>7$.
The wavelength of the Lyman-break is shifted to between $z$ band
and $J$ band for LBGs at $z=7-10$,
i.e., LBGs at $z=7-10$ have very red $z-J$ colors ($z-J > 2$, 
all the magnitudes and colors in this paper are in AB system,
unless noted otherwise).
Hence z-dropout objects are what we look for.
In the ECDFS, the deepest available $z$-band data is the HST/ACS $F850LP$ data
from the GEMS \citep[][]{rix2004,caldwell2008},
which has a $5\sigma$ limiting magnitude of 27.1.
To gain the full advantage of the deep $F850LP$ data,
$J$-band data with a $5\sigma$ limiting magnitude of 25.1
are a minimal requirement.
In the ECDFS, two $J$-band data samples are available:
the GOODS-S data taken using ISAAC on the VLT, and the MUSYC data.
The depth of the GOODS-S $J$ band is about 25.1 mag,
which meets our requirement.
The field size of the GOODS-S, however, is only about one-fifth of
that of the ECDFS,
significantly reducing the probability of finding rare and luminous LBGs
by a factor of five.
On the other hand, 
although the $J$-band data provided by the MUSYC covers the entire ECDFS,
the depth is only about 23.0 mag,
and thus far too shallow for finding $z>7$ objects.
A new deep ($J > 25.1$) near-infrared imaging survey
with the full ECDFS coverage is therefore needed for our study.

In this paper, 
we describe the deep near-infrared survey that we carried out in the ECDFS,
and the data reductions performed on the optical, near-infrared, and IRAC data,
in \S~\ref{data}.
In \S~\ref{sample}, we provide a description of the sampling criteria
for this study.
The possible contaminations in our sample are discussed in 
\S~\ref{contamination}.
We present our results and the implications of our sample in
\S~\ref{discussion}.
In \S~\ref{conclusion}, we summarize our results and discuss future work.
The cosmological parameters used in this study are
$\Omega_\Lambda = 0.73$, $\Omega_M = 0.27$, $H_0 = 71$ km s$^{-1}$ Mpc$^{-1}$,
and $w = -1$.

\section{Data and Photometry}\label{data}
\subsection{TENIS Data}
\begin{figure*}
\epsscale{1.0}
\plotone{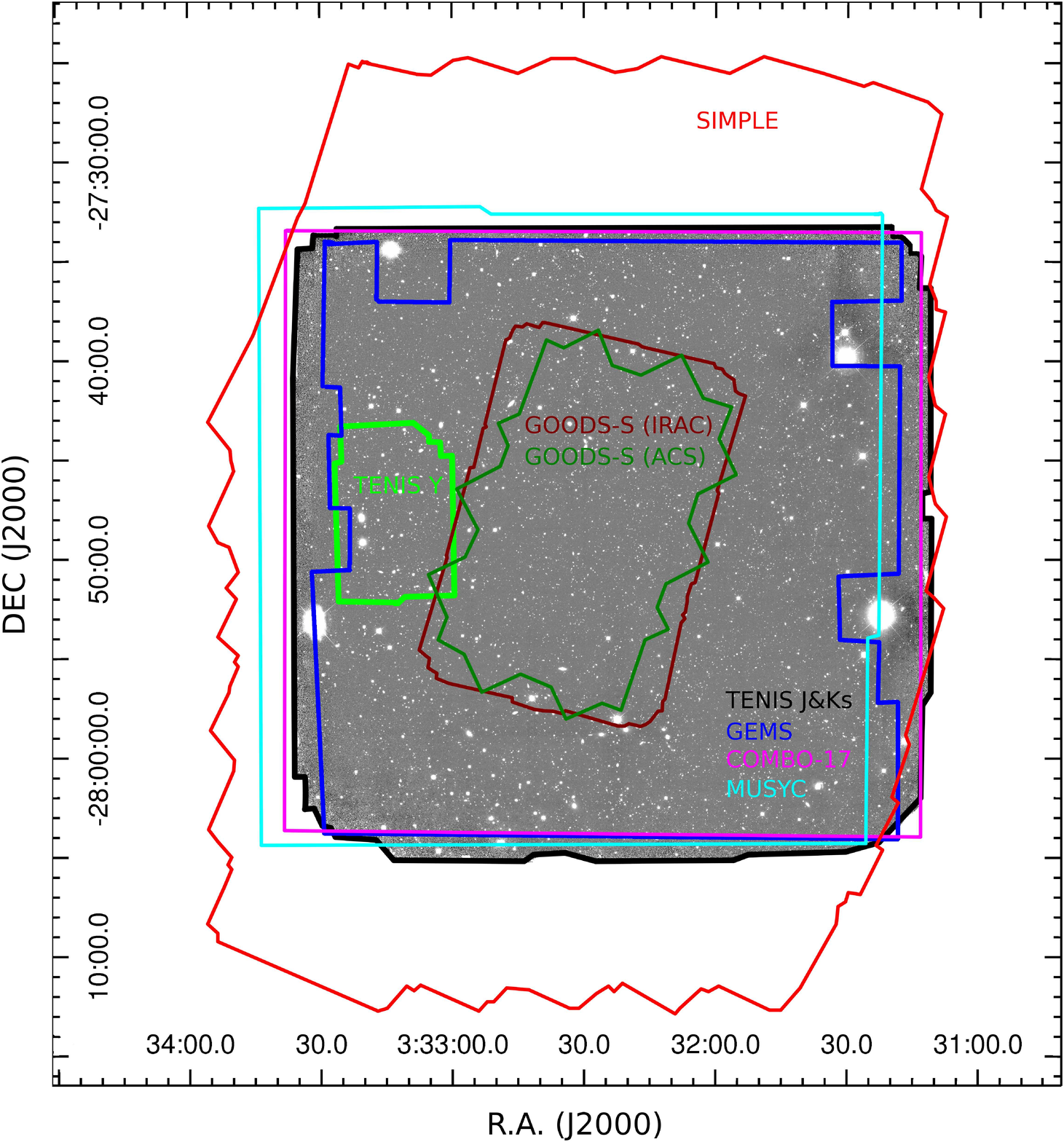}
\caption{The TENIS J-band image.
The black lines indicate the TENIS $J$ and $K_s$-band field.
The green lines delineate the TENIS $Y$-band field.
The blue, pink, cyan lines indicate the fields of GEMS, COMBO-17, and MUSYC,
respectively,
and the GOODS-S (IRAC and ACS) fields are represented by
brown and dark green lines.
\label{fov} }
\end{figure*}

We initiated the Taiwan ECDFS Near-Infrared Survey (TENIS) in 2007.
The goal of this project is to obtain deep $J$
and $K_s$ images in the ECDFS,
and thus fill in the big wavelength gap between $1.0\mu{m}$ and $3.6\mu{m}$.
TENIS is yet the deepest near-IR survey in a $30'\times30'$ sky area
with a full coverage of all four IRAC channels.
Figure~\ref{fov} shows the fields of the main surveys in the ECDFS:
SIMPLE, GOODS-S(ACS and IRAC), MUSYC, COMBO-17, GEMS, 
the TENIS $J$, $K_s$, and $Y$.
The scientific goal of the TENIS $J$-band observation
is to search for the most luminous high-$z$ objects (i.e., this work),
which are rare and thus require a survey over a large field to be found,
while the TENIS $K_s$-band observation is for the studies
of the mass assembly history and sub-millimeter galaxies at $z>2$.
After combined with existing deep images in the optical and 3.6-8$\mu{m}$,
the TENIS data can be used to derive robust photometric redshifts
and stellar masses of a large sample of galaxies.
Therefore the TENIS dataset is unique not only in its depth and survey area
but also in its rich multi-wavelength ancillary data in the ECDFS.

\subsubsection{CFHT WIRCam Observations}
The TENIS data were taken using the Wide-field InfraRed Camera \citep[WIRCam,][]{puget2004} on 
the Canada-France-Hawaii Telescope (CFHT).
WIRCam consists of four 2048$\times$2048 HAWAII2-RG detectors
covering field of view of $20'\times20'$ with a $0''.3$ pixel scale.
To date we have spent 44 hours in $J$ and 40 hours in $K_s$.
The $J$-band data were taken in 2007B and 2008B, 
with an average $5\sigma$ limiting magnitude 
of $J=25.5$ for point sources.
The $K_s$-band data were taken in 2009B and 2010B, 
with an average $5\sigma$ limiting magnitude of $K_s=25.0$.
The average seeings were $0''.8$ and $0''.7$ for $J$ and $K_s$, respectively.
More details of the WIRCam imaging observations
is described in the TENIS catalogue paper (Hsieh et al., in preparation).

\subsubsection{Subaru MOIRCS Observation}\label{moircs}
We obtained $Y$ data in the ECDFS using the Subaru MOIRCS in 09B.
The observation is a follow-up observation of the TENIS project.
The goal of the observation is to have $Y$-band photometry
for a set of high-$z$ objects pre-selected using their $z - J$ colors.
Two pointings were planned to be observed.
One is inside the GOODS-S field and the other is outside the GOODS-S field.
However, due to the bad weather, only one pointing on the east side of the ECDFS was observed,
i.e., a $4'\times7'$ area outside the GOODS-S field.
The $5\sigma$ limiting magnitude for point sources is 25.5.
Although the $Y$ data are only available for part of the ECDFS,
the data are still very important to further improve the photometric
redshift quality for the candidates in that area
since the wavelength of $Y$ band is right between those of $F850LP$ and $J$.

\subsubsection{Data Reduction}\label{datareduction}
The TENIS $J$ and $K_s$ data were processed using 
an Interactive Data Language (IDL) based reduction pipeline 
called Simple Imaging and Mosaicking Pipeline.
The details of the pipeline are described in \citet{wang2010}.
We used the pipeline to deal with flat-fielding, 
removing instrumental features like crosstalk and residual images 
from saturated objects in previous exposures,
masking satellite trails, distortion and astrometry correction, mosaicking and
stacking images, and photometric calibration. 
The zero-point calibration was done by matching the fluxes 
in an aperture of $5''$ in diameter with those in 
the Two Micron All Sky Survey (2MASS) point source catalogs.

\subsubsection{Photometry}\label{photometry}
We used SExtractor version 2.5.0 \citep[][]{ba1996}
to detect objects and measure their fluxes in the WIRCam images. 
For the TENIS $J$ image, 
we used the ``FLUX\_AUTO'' values of the SExtractor output 
as the measured $J$ fluxes.
The flux errors provided by SExtractor do not include the correlated errors.
Therefore we calibrated the errors with the following procedure:
First, the fluxes and the flux errors were re-measured 
using SExtractor with a $2''$ diameter aperture.
We then convolved the source-masked image with a $2''$ diameter aperture,
and calculated the rms around each pixel on the convolved images.
The ratio between the aperture photometric flux error provided by SExtractor 
for a certain object and the rms value 
around the same position in the convolved image
is the correction factor of the flux error for that object.
The median value of the correction factors was computed to be 
the general correction factor for all the sources.
For $J$-band data, the general correction factor is 0.793.
The same procedures were done with different aperture sizes.
We find that the correction factors are very stable 
with different aperture sizes,
which is consistent with the experience in Wang et al. 2010.
Therefore we just applied the factor to the ``FLUXERR\_AUTO'' values 
for all the TENIS $J$ objects to calibrate their flux errors.

For the TENIS $K_s$ image, the double-image mode of SExtractor 
was performed to generate a $J$-selected $K_s$ photometric catalogue.
In order to use the double-image mode, 
we shifted the $K_s$ image to match to the $J$ image.
The $K_s$ fluxes were also measured with the ``FLUX\_AUTO'' option,
and the flux errors were calibrated with the same procedure used in the
TENIS $J$ data.

\begin{figure}
\epsscale{1.0}
\plotone{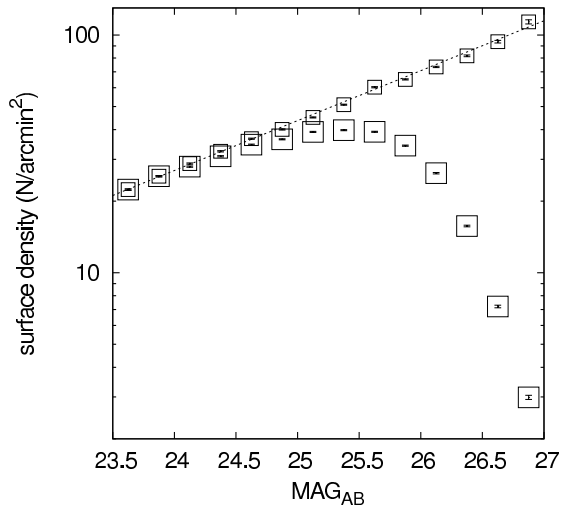}
\caption{The surface density vs. TENIS $J$-band magnitude plot.
The surface density is measured by selecting sources using SExtractor.
The large open box indicates the surface density for each magnitude bin.
The small open box indicates the surface density for each magnitude bin
after the completeness correction.
The dotted line is the best fit of the surface density 
after the completeness correction.
\label{completeness} }
\end{figure}

To estimate the completeness for our data, we used a Monte-Carlo simulation.
The faint objects are compact but may not be all point-like.
The medium full width half maximum (FWHMs) of objects 
with $J\sim25$ is about $2''.0$ ($\sim10$ kpc at $z\sim7$).
Therefore, to better determine the completeness of faint objects,
we randomly put 100 artificial sources into the TENIS $J$ image,
with an image that is averaged from $\sim100$ real extended sources 
with FWHMs between $2''.0$ and $2''.2$, 
and with signal-to-noise ratios greater than 100.
The magnitudes of the artificial sources are uniformly distributed
in the magnitude space from 20mag to 27mag,
and Poisson errors were added into the images of the artificial sources.
We then ran SExtractor to do object finding.
This procedure was run for 1000 times to deliver the detection rate
of these artificial sources in each magnitude bin, i.e., the completeness.
Figure~\ref{completeness} shows that the surface density vs. apparent magnitude
of the TENIS $J$ data.
The completeness corrected surface density is used in our analyses later.

\subsection{Ancillary Data}
We incorporated data from several ECDFS surveys at different wavelengths
to perform our analyses.
To measure accurate colors for the color selection,
we processed the images from the ancillary data
to generate new images that are pixel-to-pixel matched to the TENIS images.
For the space-based optical images, we also matched their PSFs to the TENIS data
to ensure better color measurements.
In this subsection, we describe how we processed the ancillary data
and measured fluxes.

\subsubsection{MUSYC Data}\label{musyc}
The MUSYC team has released their deep optical 
and near-infrared images in the ECDFS.
To improve the quality of the photometric redshifts,
which will help eliminate foreground contaminations in our study,
we included the MUSYC optical data in our analyses.
We resampled the MUSYC images in $U$, $U38$, $B$, $V$, $R$, $I$, $z$, and $O_3$
to match the TENIS $J$ image, 
and then used the double-image mode of SExtractor with ``FLUX\_AUTO'' 
to generate a TENIS $J$-selected MUSYC photometric catalogue.
The zero-point calibration was done by matching the fluxes with those 
in the published $BVR$-selected MUSYC catalogue \citep[][]{gawiser2006}.
The flux errors of the MUSYC photometry were calibrated 
using the method described in \S~\ref{photometry}.

\subsubsection{GEMS ACS Data}\label{acs}
The GEMS project \citep[][]{rix2004,caldwell2008} provides deep HST ACS images
with the $F606W$ and $F850LP$ filters in the ECDFS,
except some small areas close to bright objects.
The $5\sigma$ limiting magnitudes for point sources are
28.4 and 27.1 for $F606W$ and $F850LP$, respectively.
We mosaicked the public GEMS ACS images
and then resampled the images to match the TENIS $J$ image.
The PSFs of the ACS images 
are very small compared to the PSFs of the WIRCam images.
To derive better color measurements between the ACS filters and $J$,
we applied a smoothing kernel to the ACS images to match their PSFs
to that of the TENIS $J$ image.
We then repeated the same procedure as what we did for the MUSYC data
to deliver the $J$-selected GEMS photometric catalogue 
with calibrated flux errors.

\subsubsection{SIMPLE IRAC Data}\label{deconvolve}
The SIMPLE project \citep[][]{damen2011} provides deep
IRAC observations covering the entire ECDFS
with the $10'\times15'$ GOODS-S IRAC mosaics 
in its center \citet{dickinson2003}.
We mosaicked the SIMPLE images and resampled them to match the TENIS $J$ image.
For the flux measurement, however, we could not use SExtractor in the
double-image mode with the ``FLUX\_AUTO'' option
to generate the $J$-selected IRAC photometric catalogue,
since the PSFs of IRAC images are about $1''.5$ to $2''.0$ arcsec,
which are much larger than that of the TENIS $J$ image.
If we were to do this, the IRAC fluxes would be seriously under-estimated
as SExtractor uses the aperture size estimated
from the reference image (TENIS $J$ image) to measure the fluxes
on the target image (SIMPLE IRAC images).

One can use SExtractor in the double-image mode with
the standard aperture photometry to measure the IRAC fluxes,
and then apply the aperture correction values provided
in the IRAC data handbook (Chapter 5, Table 5.7) to
derive the estimated IRAC total fluxes.
However, due to the very large PSFs of IRAC images,
using simple aperture photometry to measure the fluxes
of deep IRAC images in a relatively crowded field like the ECDFS
has serious confusion issues
and the local background estimation used in aperture photometry
can be easily affected by neighboring objects.
We note that the accuracy of the IRAC photometry is critical
in our analyses (see \S~\ref{sample}),
requiring an alternative method to estimate the IRAC fluxes properly.

Many intensive efforts have gone into developing methods for estimating fluxes
in an image with a large PSF (image1, hereafter).
The most recent method relies on utilizing the morphological information
from a high quality image in another band (image2, hereafter),
\citep[e.g.,][]{grazian2006,laidler2007,wang2010,mclure2011}.
By assuming the intrinsic morphology of objects are identical
in the two wavelengths (i.e., no color gradient from its center to edge),
one is able to use the PSF of image2 to deconvolve that image,
and then convolve the deconvolved image2 with the PSF of image1.
The scaling factor that needs to be applied to match the peak values 
between an object in image2 and the same object in the processed image1
is the flux ratio (color) of that object.
The assumption that there is no color gradient, however, does not always stand.
We have therefore developed a new method 
that is quite different from what has been used, 
but is very similar to the traditional CLEAN deconvolution in radio imaging.

We used the TENIS $J$ image as a prior image (i.e., image2)
to estimate the fluxes in the IRAC images (i.e., image1).
The segmentation map of the TENIS $J$ image generated by SExtractor with
a $0.7\sigma$ threshold preserves the information of 
what pixels are attributed to each object,
and it provides the positional and shape/boundary information of each object
when deconvolving the IRAC images.
We only used the boundary information rather than 
the full morphological information adopted in previous works.
The boundary information introduced with the segmentation map 
is often called the ``CLEAN window'' in radio image.
The IRAC PSFs were generated using bright isolated IRAC point sources
in each IRAC channel.
The size of the PSF images is $1'\times1'$.
Due to the non-spherical IRAC PSF and different orientations of $Spitzer$
in the observing epochs for GOODS-S and SIMPLE,
the PSFs of the stacked IRAC image in GOODS-S region is very different from
those in the extended area.
We tried to use the PSF generated from the SIMPLE data 
to deconvolve the GOODS-S region 
and found that the fluctuation of the residual image 
(see the next paragraph for the detail of the residual image) is 5 times larger
(i.e., $\sim1.7$ mag worse in limiting magnitude) 
as compared to the result using the PSF generated from the GOODS-S data itself.
We therefore generated an individual PSF for each region
and deconvolved the two regions separatedly.

The deconvolution process always starts at the pixel with the highest value
measured within a 9 pixel aperture in diameter ($F_{AP9}$, hereafter),
and this pixel must have been registered to one object in the
$J$ segmentation map.
Since the SIMPLE image ($\sim0''.6$ per pixel) has been resampled 
to match the TENIS $J$-band image ($\sim0''.3$ per pixel),
simply moving a window (i.e., $F_{AP9}$) across the SIMPLE image 
to find the location of a peak is very similar to do a sub-pixel centering.
Once this pixel is found,
we subtracted a scaled PSF from the surrounding $1'\times1'$ area 
centered on this pixel.
The ratio of the subtracted flux depends on $F_{AP9}$;
if $F_{AP9}$ is greater than $5\sigma$, then 0.1\% of the flux was subtracted.
If $F_{AP9}$ is less than $5\sigma$, then 100\% of the flux was subtracted.
The subtraction ratio that we use (0.1\%) is much smaller than 
the values that most deconvolution/clean algorithms usually use (10\%).
We found that a subtraction ratio of 10\% for our deconvolution process
sometimes produces much worse results especially for bright extended sources
because our algorithm is less restricted;
0.1\% is the value that has a good balance between performance and quality.
The subtracted flux ($F_{SUB}$) was summed and registered to 
a certain reference object according to the $J$ segmentation map.
After the subtraction, 
the deconvolution process is repeated on the subtracted image,
until there was no pixel with $F_{AP9}$ higher than $1.5\sigma$.
We note that the concept of the deconvolution here 
is identical to that of CLEAN in radio imaging.
At the end of the deconvolution process,
the $F_{SUB}$ for each object is the flux measurement of that object,
and the final subtracted image is the residual map.
The latter allows us to check the quality of the deconvolution,
and it is also used for estimating accurate photometric errors.
The flux error of each object was calcualted based on 
the fluctuation of the local area around that object in the residual map.
Any imperfection of the PSF would cause larger fluctuation in the
residual map,
and that was taken into account by the flux error calculation.

\begin{figure*}
\epsscale{1.0}
\plotone{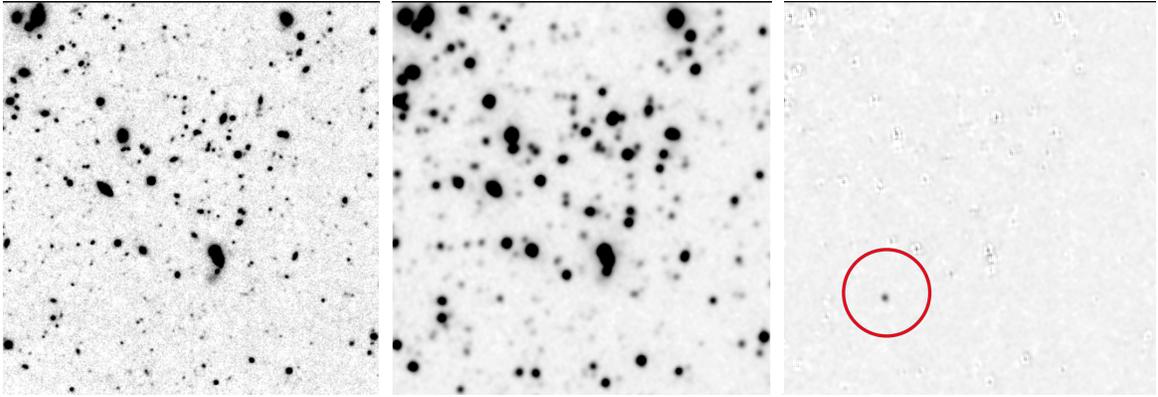}
\caption{
These three panels show the deconvolution quality of the IRAC images.
A small region of the TENIS $J$ image, the SIMPLE $3.6\mu{m}$ image,
and the residual SIMPLE $3.6\mu{m}$ image are shown in the
left, middle, and right panels.
The region of each panel is $135''\times135''$.
North is up and east is to the left.
The brightness and contrast of the three panels are exactly the same.
The clean residual image in the right panel
shows that the deconvolution works very well.
The unsubtracted object marked by the red circle in the right panel
is an IRAC-bright source but that is not detected in $J$.
\label{irac-decon} }
\end{figure*}

We demonstrate the performance of this method in Figure~\ref{irac-decon}.
A small region of the TENIS $J$ image, the SIMPLE $3.6\mu{m}$ image,
and the residual SIMPLE $3.6\mu{m}$ image are shown in the
left, middle, and right panels.
The brightness and contrast of the three panels are exactly the same.
By comparing the SIMPLE $3.6\mu{m}$ image to the TENIS $J$ image,
one can see that many objects in the most dense area are mixed together
in the $3.6\mu{m}$ image while they are still well-separated in the
TENIS $J$ image.
After processing the $3.6\mu{m}$ image using the method described above,
the residual image
as shown in the right panel of Figure~\ref{irac-decon}
shows that the deconvolution works very well.
Sources left in the residual image are objects 
detected in the IRAC images but not the TENIS $J$ image.
An example is the object sitting at the center of the red circle 
in the right panel of Figure~\ref{irac-decon}.
This is a bonus of the deconvolution method.

\begin{figure}
\epsscale{1.2}
\plotone{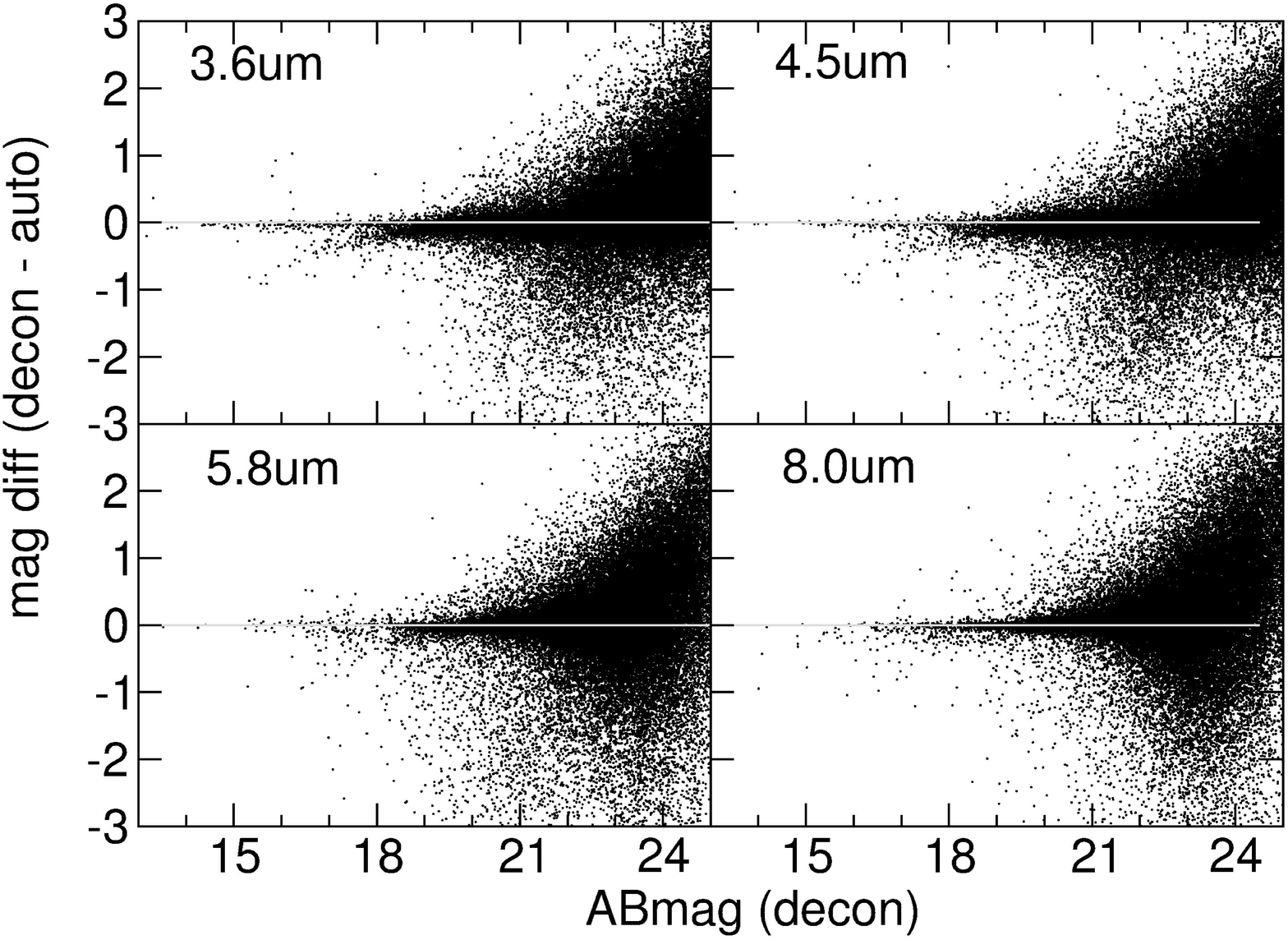}
\caption{The comparison of the IRAC photometry using difference methods.
The differences between the magnitudes using the deconvolution method (decon)
and that using the SExtractor with FLUX\_AUTO (auto) are shown.
It shows that the photometry using two different methods are consistent
for bright objects
but the scatter between the two measurements increases for fainter objects.
\label{irac-comp} }
\end{figure}

Figure~\ref{irac-comp} shows a comparison of 
IRAC photometry using the abovementioned deconvolution method against
that using the SExtractor with FLUX\_AUTO.
As can be seen,
the results of two photometric methods 
agree with each other for bright objects, 
but the scatter between the two measurements increases for fainter objects.
This is because the flux measurements of faint objects 
are much more easily affected by neighboring objects,
coupled with the fact that 
fainter objects have higher mean surface number densities.
Some bright sources have $>0.1$ mag differences 
between their deconvolution and auto magnitudes.
According to the flags provided by SExtractor,
more than 80\% of objects brighter than 20 mag are seriously blended
with their neighbors.
We also confirmed that all the objects having large differences
between their deconvolution and auto magnitudes are blended
with their neighbors as shown in their flags.
The deconvolution method can deliver more accurate flux
estimation for faint sources in the IRAC images,
an important factor in our selection criterion for $z>7$ objects
(see \S~\ref{sample}).
A more detailed description of the deconvolution method we used
to estimate the IRAC fluxes is described in Hsieh et al. (in preparation).

\subsection{$J$-selected master catalogue}\label{catalogue}
We combined all the data described in this section
and generated a 16-band $J$-selected master catalogue containing 114K objects.
The $5\sigma$ limiting magnitude for point sources in each band is shown 
in Table~\ref{limiting-mag}.
For the MUSYC, GEMS, and TENIS data,
the values were derived using an $1''.0$ aperture in diameter 
with aperture and noise correlation correction.
According to \citet{rix2004} and \citet{caldwell2008},
the $5\sigma$ limiting magnitudes for the GEMS $F606W$ and $F850LP$ 
are 28.53 and 27.27, 
which are much deeper than those listed in Table~\ref{limiting-mag},
because the aperture size that they use is $\sim0''.1$ in diameter,
which is 10 times smaller as compared to ours.
For the SIMPLE data, 
the aperture we used in the IRAC deconvolution process is a 9-pixel window,
which mimics an $3''.0$ aperture in diameter for point sources.
The IRAC limiting magnitudes in Table~\ref{limiting-mag},
which are the medium values derived using point sources 
with $5\sigma$ detections in both SIMPLE and GOODS-S IRAC data.
The values are different from those listed in \citet{damen2011}
because of the different methodologies.
More details are provided in the TENIS catalogue paper
(Hsieh et al., in preparation).
This master catalogue is used to select our $z>7$ sample.

\begin{deluxetable}{lccc}
\tablecolumns{2}
\tablewidth{0pt}
\tablecaption{$5\sigma$ limiting magnitudes for point sources 
in the $J$-selected master catalogue\label{limiting-mag}}
\tablehead{
\colhead{Band} & \colhead{$5\sigma$ limiting magnitude}}
\startdata
MUSYC U    &  26.44\tablenotemark{a}  \\
MUSYC U38  &  26.34\tablenotemark{a}  \\
MUSYC B    &  27.21\tablenotemark{a}  \\
MUSYC V    &  26.85\tablenotemark{a}  \\
MUSYC R    &  26.80\tablenotemark{a}  \\
MUSYC I    &  25.06\tablenotemark{a}  \\
MUSYC z    &  24.29\tablenotemark{a}  \\
MUSYC O3   &  25.88\tablenotemark{a}  \\
GEMS ACS F606W & 26.80\tablenotemark{b} \\
GEMS ACS F850LP & 25.72\tablenotemark{b} \\
TENIS J    &  25.57  \\
TENIS Ks   &  24.95  \\
IRAC $3.6\mu{m}$ &  25.82 \\
IRAC $4.5\mu{m}$ &  25.75 \\
IRAC $5.8\mu{m}$ &  23.77 \\
IRAC $8.0\mu{m}$ &  23.62 \\
\enddata
\tablenotetext{a}{The MUSYC limiting magnitudes in this table are deeper than
those in \citet{gawiser2006} and \citet{taylor2009} 
because we use a smaller aperture size.}
\tablenotetext{b}{The GEMS limiting magnitudes in this table are shallower
than those in \citet{rix2004} and \citet{caldwell2008}
because we use a much larger aperture size. See text for details.}
\end{deluxetable}

\section{CANDIDATE SELECTION}\label{sample}

\subsection{Color criterion}\label{colorcriterion}
For $z$-dropout studies, 
the most important issue is to eliminate contaminations from
low-$z$ galaxies and Galactic cool stars.
Most previous studies make great efforts to estimate the contamination rates,
which are not negligible.
The accuracies and uncertainties of these estimates, however, are unknown;
the best way to deal with the contaminations is to eliminate them
in the first place as far as is possible.
The IRAC photometry provide a good solution to this issue
and we discuss how to utilize the IRAC photometry to select
a relatively pure sample in this subsection.

\begin{figure}
\epsscale{1.1}
\plotone{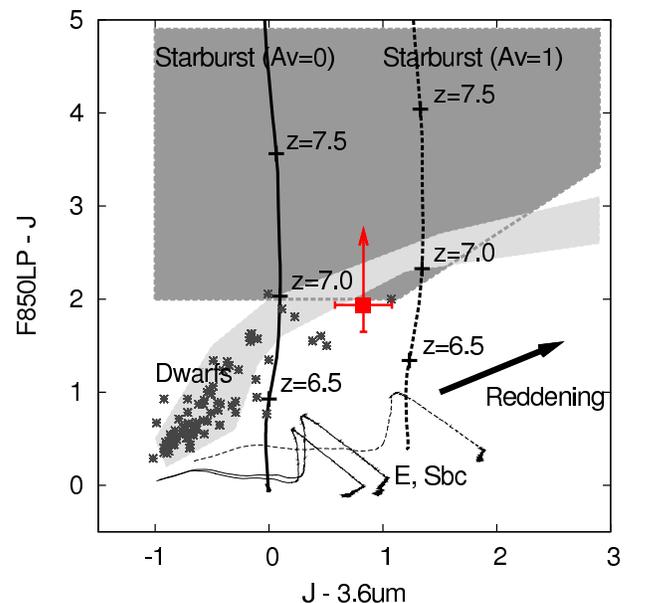}
\caption{Selection criterion in $J-3.6\mu{m}$ vs. $F850LP - J$ diagram.
The thick solid line indicates LBGs at $z>5.0$
and the thick dotted line indicates LBGs with $Av=1$ at $z>5.0$
The two thin solid lines indicates early-type and late-type galaxies
at $z<5.0$,
and the thin dotted line indicates late-type galaxies with $Av=1$ at $z<5.0$.
The black arrow shows the reddening direction.
The dark gray area represents the selection criteria of Equation~\ref{criteria1}
while the light gray area indicates where the Galactic cool stars are,
which are estimated using the AMES-dusty model \citep[][]{allard2001}.
The observed M and L dwarfs from IRTF Library are plotted with dark star signs.
The red dot indicates the candidate that we find (see \S~\ref{contamination} for the details).
This figure shows that although the selection criteria can pick up LBGs
at $z>7$ and avoid low-$z$ galaxies with any level of reddening,
some L dwarf stars could still be selected as well.  
\label{cc-diag} }
\end{figure}

Galaxies at different redshifts in a $J-3.6\mu$m vs. $F850LP - J$ diagram
are shown in Figure~\ref{cc-diag}.
The GALAXEV template \citep[][]{bc2003} is used to generate this plot.
LBGs at $z>5$, late-type, and early-type galaxies at $z<5$ with different
extinctions are plotted.
Most of the previous $z$-dropout studies use a simple color cut
(e.g., $z - J > 2.0$) to select $z>7$ objects.
However, as shown in Figure~\ref{cc-diag}, heavily reddened low-$z$ galaxies
would have very red $z - J$ colors, 
and the $z - J$ color of a late-type galaxy with $Av > 3$ would be 
very similar to that of $z>7$ objects ($z - J > 2$).
Our two-color criteria in the $J-3.6\mu$m and $F850LP - J$ space,
however, greatly minimizes this issue.  
The selection criterion are accordingly
\begin{eqnarray}
F850LP-J > 2.0,  \nonumber \\
J-3.6\mu{m} > -1.0,  \nonumber \\
F850LP-J > 0.8\times(J-3.6\mu{m}) + 1.1,
\label{criteria1}
\end{eqnarray}
as shown by the dark gray area in Figure~\ref{cc-diag}.
The third criterion is closely parallel to the reddening direction
of low-$z$ galaxies,
and therefore greatly reduces the contamination rate from low-$z$ galaxies.  

Although contamination by low-$z$ galaxies can be effectly dealt with 
using the criteria defined in Equation~\ref{criteria1},
contamination by Galactic cool stars can still be problematic.
We use the AMES-dusty dwarf star template \citep[][]{allard2001} to estimate
the distribution of M, L, and T dwarfs in Figure~\ref{cc-diag},
which is plotted as a light gray area.
The observed M and L dwarfs from IRTF Library 
\footnote{\url{http://irtfweb.ifa.hawaii.edu/$\sim$spex/IRTF\_Spectral\_Library/}}
are also shown as dark star signs. 
We note that there is an overlapped region between our selection criteria
(dark gray area) and the dwarf distribution (light gray area).
According to the AMES-dusty template, 
dwarfs in the overlapped region are L dwarfs with effective temperatures 
between 1700K and 1900K.
Therefore, we need to make another criterion 
to discriminate against very cool dwarf stars.

\begin{figure}
\epsscale{1.1}
\plotone{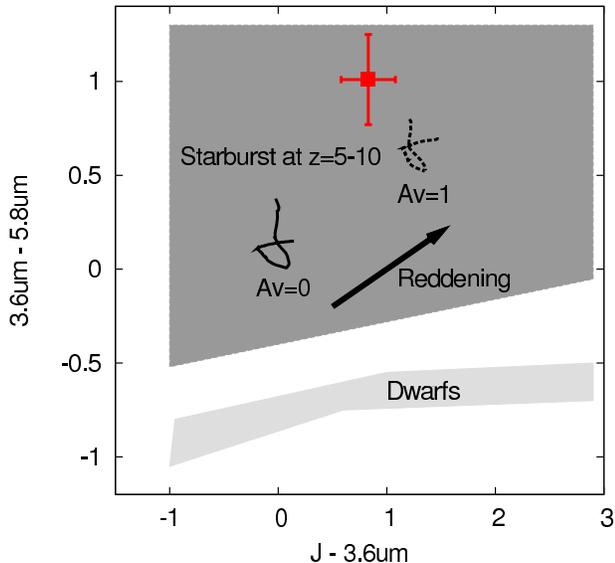}
\caption{IRAC color-color diagram for high-$z$ galaxies and dwarf stars.
The solid line and dashed line indicate starburst galaxies at $z=5-10$
with $A_V=0$ and $A_V=1$, respectively.
The arrow indicates the reddening direction.
The dark gray area shows the selection criteria for high-$z$ galaxies
while the light gray area shows where the dwarf stars are.
The red dot indicates the candidate that we find (see \S~\ref{contamination} for the details).
According to this figure, using IRAC colors can successfully break
the color degeneracy between high-$z$ galaxies and dwarfs in optical
and near infrared bands.
\label{cc-diag-irac} }
\end{figure}

Fortunately, the IRAC colors of dwarf stars are very different from
those of high-$z$ galaxies.
Therefore we can minimize the contamination issue from Galactic cool stars
by adding another selection criterion:
\begin{eqnarray}
3.6\mu{m}-5.8\mu{m} > 0.12\times(J-3.6\mu{m})-0.4.
\label{criteria2}
\end{eqnarray}
Figure~\ref{cc-diag-irac} shows how this additional criterion works.
The solid and dashed lines indicate starburst galaxies at $z=5-10$
with $A_V=0$ and $A_V=1$, respectively.
The arrow indicates the reddening direction.
The dark gray area shows the selection criteria for high-$z$ galaxies
while the light gray area shows the region occupied by dwarf stars
as computed using the AMES-dusty \citep[][]{allard2001} model.
In Figure~\ref{cc-diag-irac}, 
high-$z$ galaxies and dwarf stars are well separated,
demonstrating that IRAC colors can break the color
degeneracy between high-$z$ galaxies and dwarf stars
in the optical and near-infrared bands.

To improve the quality of the candidate selection, 
additional criteria are added
to eliminate low-$z$ galaxies that satisfy the above selection criterion
because of photometric errors. 
First, all candidates must have more 
than $3\sigma$ detection in the WIRCam $J$ band. 
Second, all candidates must have fluxes lower than $1.5\sigma$ 
in optical bands bluer than $F850LP$.
We also stack all the optical images (except the $F850LP$ and MUSYC $z$ bands
as objects at $z>7$ may still be detected in $z$ bands) 
to generate an ultra-deep optical image.
\citet{mclure2011} have found that some previously published $z$-dropout
candidates actually have counterparts in ultra-deep stacked optical images,
that are just contaminations.
The ultra-deep stacked optical image therfore allows us to check 
if some contaminations cannot be detected in our single optical bands
because of insufficient sensitivity. 
We generate two different stacked optical images:
one is averaged from MUSYC $U, U38, B, V, R, I,$ 
and GEMS ACS $F606W$ (7 filters); 
the other is averaged from just red filters 
(MUSYC $V, R, I,$ and GEMS ACS $F606W$).
The former provides the deepest stacked optical image
while the latter is optimized for detecting faint red objects.
At this stage, 143 objects satisfy all the color criteria.

\subsection{Photometric redshift criterion}\label{photoz}
Although the two color-color selection criteria are very robust for
picking up $z>7$ objects,
contaminations can still be present
because of photometric errors and transients.
Photometric redshift can help to further reduce these issues.
It utilizes all the data (16 bands) rather than just
the 4 filters used in our color-color selections,
so it is less affected by photometric errors and transients.
We use the EAZY code \citep[][]{brammer2008} to estimate 
the photometric redshift of our candidates.
We allow a $5\%$ systematic flux error for each band
to take into account the systematic errors caused by the fact 
that for a given object we are not missing the same fraction of light 
in different passbands that have very different spatial resolutions. 
To minimize the bias effect of the photometric redshift estimation
due to different fitting templates, 
we use three different templates to run the EAZY code,
thus obtaining three photometric redshifts for each candidate.
The three templates used are CWW+KIN, default EAZY v1.0, and P\'{E}GASE.
The CWW+KIN template is the CWW empirical template \citep[][]{cww1980} 
with the extension prescribed by \citet{kinney1996}.
The default EAZY v1.0 template is generated using
the \citet{br2007} algorithm with the P\'{E}GASE models 
and calibrated using semi-analytic models,
plus an additional young and dusty template.
The P\'{E}GASE template is from \citet{fr1997}.

\citet{yan2004} have found that many IRAC-selected extremely red objects (IERO)
with very high photometric redshifts are actually galaxies
at $z<2$ possessing two different stellar populations.
The EAZY code can fit data
with multiple stellar populations simultaneously.
For an IERO or similar object, using this function helps to break 
the degeneracy of high and low redshifts in the template fitting.
We use this multiple population fitting function when employing
the CWW+KIN and default EAZY v1.0 templates.
The documentation on the P\'{E}GASE template suggests that
this template works better in the single population fitting mode,
and so we decided not to use the multiple population fitting function
with the P\'{E}GASE template.

We compile a new list of candidates according to their photometric redshifts.
We pick up an object with a photometric redshift between 6.0 and 9.0,
and the lower-limit of its photometric redshift at 68\% confidence level
must be greater than 3.0.
Since there are three different photometric redshifts for each object,
all three photometric redshifts for a chosen candidate 
must meet the selection criterion simultaneously.
We then assign the photometric redshift with the smallest best-fitted chi-square
as the redshift of each candidate.
Applying the photometric redshift criterion 
left 22 objects in our candidate list.

\subsection{Cleaning the Sample}\label{purify}
All the previous selection criteria rely entirely on 
photometry (including the photometric redshift estimation).
Photometric measurements, however, 
can be easily affected by CCD bleeding, satellite tracks,
diffraction spikes of bright stars, etc., 
which will also affect the performance of the selection criteria we use.
To eliminate these false-detection cases, 
we visually check the images of every bands for 
all the 22 remaining candidates and remove 2 affected objects from the list.
In addition, objects detected in $24\mu{m}$, X-ray, and 1.4GHz are also omitted 
since they are very unlikely to be at high $z$.
There are 3 objects rejected at this stage.

Although we use a color selection criterion
in the $J - 3.6\mu{m}$ vs. $3.6\mu{m} - 5.8\mu{m}$ space 
to discriminate against Galactic cool stars,
this criterion does not work well for some objects 
which are very faint in $3.6\mu{m}$ and $5.8\mu{m}$.
This increases the contamination rate in the candidate selection
because the uncertainties in their colors 
make them spill over into the selection region.
To deal with this issue, we perform the SED fitting 
for the candidates using the AMES-dusty dwarf star template.
We reason that an SED fitting would provide better information 
about the spectral type of an object
as it uses the photometry from all the bands 
as compared to a pure color selection.
We reject objects if they have similar or even smaller minimum chi squares 
with the AMES-dusty template as compared to those with the galaxy templates.
18 objects are rejected in this way.

\begin{figure}
\epsscale{1.0}
\plotone{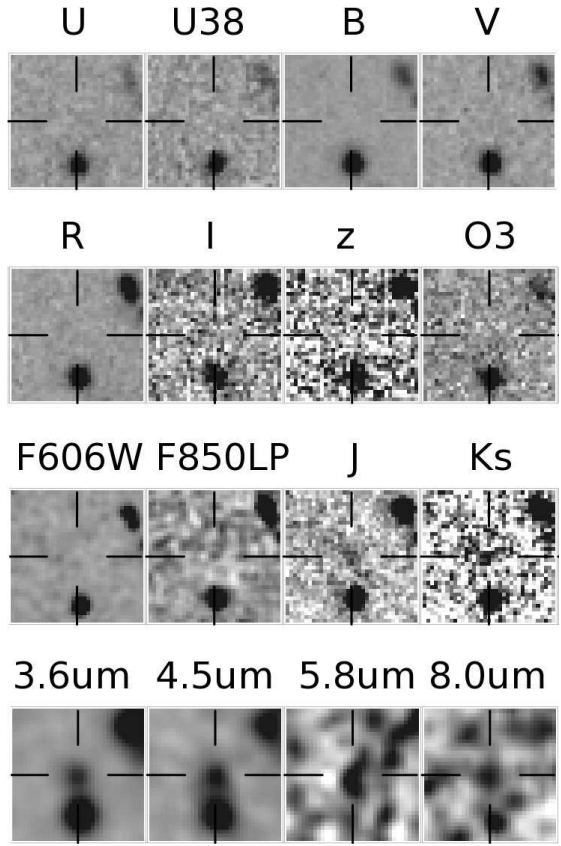}
\caption{Thumbnail images for TENIS-ZD1.
The region of each image is $10''\times10''$.
North is up and east is to the left.
TENIS-ZD1 is not detected in all the optical bands
and detected in all the WIRCam and IRAC images.
\label{thumbnail} }
\end{figure}

\begin{figure}
\epsscale{1.0}
\plotone{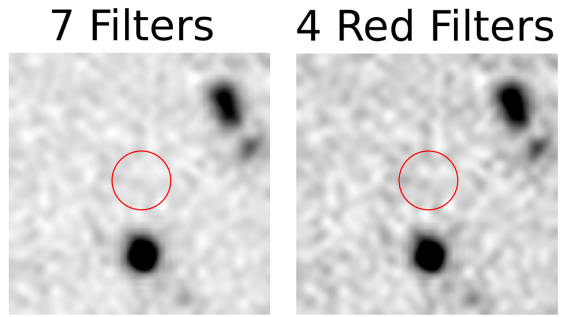}
\caption{Ultra-deep stacked optical images for TENIS-ZD1.
The left-hand image is averaged using MUSYC
$U, U38, B, V, R, I,$ and GEMS ACS $F606W$ images (7 filters).
The right-hand image is combined from
MUSYC $V, R, I,$ and GEMS ACS $F606W$ images (4 red filters).
The region of each image is $12''\times12''$.
North is up and east is to the left.
The details of the ultra-deep stacked optical images are in \S~\ref{purify}.
The red circle indicates the location of TENIS-ZD1 in the $J$ image.
According to these two images, 
the optical flux of TENIS-ZD1 is below the detection limit
even in ultra-deep optical images.
\label{opt-comb} }
\end{figure}

\begin{figure*}
\epsscale{1.0}
\plotone{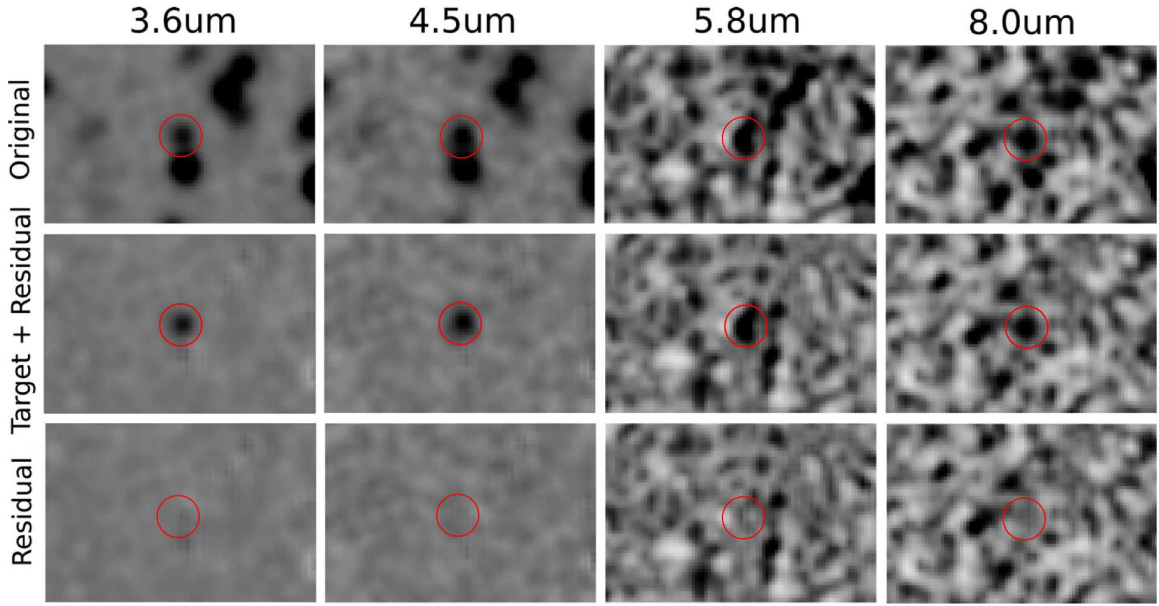}
\caption{The IRAC residual images of TENIS-ZD1.
Images from left to right indicate images in IRAC $3.6\mu{m}$, $4.5\mu{m}$,
$5.8\mu{m}$, and $8.0\mu{m}$.
The upper images are the original images of IRAC channels.
The middle ones are the images 
after subtracting all the neighbors (except the target) detected in $J$ 
using the deconvolution method.
The lower ones are the residual images 
after subtracting all the objects detected in $J$.
The red circle indicates the location of TENIS-ZD1 in the $J$ image.
The region of each image is $25''\times15''$.
North is up and east is to the left.
By comparing these three images,
they show that TENIS-ZD1 and its bright close neighbor
are well-deconvolved and well-deblended,
which suggests that the IRAC flux measurement effect of TENIS-ZD1
due to the bright close neighbor is minimized.
\label{41126-deconv} }
\end{figure*}

\begin{figure}
\epsscale{1.1}
\plotone{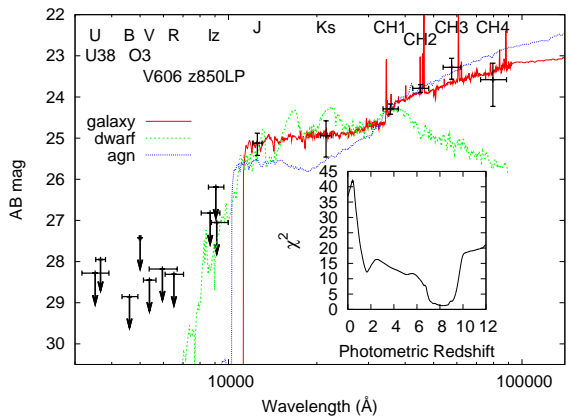}
\caption{The best-fitted SED and the $\chi^2$ vs. redshift plot.
The black data points indicate the AB magnitudes of TENIS-ZD1 in 16 bands.
The red, green, and blue lines indicate the best-fitted templates of
galaxy, dwarf star, and AGN/quasar, respectively.
The $\chi^2$ plot for the best-fitted galaxy template 
shows that the photometric redshift fitting of TENIS-ZD1 
has a very low $\chi^2$ from $z_{phot}=7.0$ to $z_{phot}=9.0$,
while the weighted $z_{phot}$ is 7.8.
\label{sed} }
\end{figure}

In the end, only one object survives all the criteria imposed.
We name this object TENIS-ZD1.
We show its thumbnails in Figure~\ref{thumbnail}
and summarize its photometric results in Table~\ref{fluxes}.
The listed detection limits for the optical bands are $1\sigma$ values,
and the size of the Kron aperture used for TENIS-ZD1
is $3''.2\times1''.7$ (major and minor axes) 
according to the output of SExtractor.
The FWHM of TENIS-ZD1 in $J$ provided by a Gaussian fitting is $2''.3\pm0''.46$,
demonstrating that it is an extended object (and hence not a star).
The ultra-deep stacked optical images are also shown in Figure~\ref{opt-comb}
to show that TENIS-ZD1 is not detected in both ultra-deep optical images.
As we mentioned in \S~\ref{deconvolve}, 
accurate IRAC flux measurements are essential for picking up our candidates.
Figure~\ref{thumbnail} shows that TENIS-ZD1 has a bright close neighbor
with a separation of just $3''$.
Therefore the IRAC photometry of TENIS-ZD1 measured using a typical
aperture photometry method would be seriously affected by this bright neighbor.
In \S~\ref{deconvolve} we describe the deconvolution method 
that we use to derive IRAC fluxes,
and Figure~\ref{41126-deconv} 
shows a comparison between the original IRAC images around TENIS-ZD1,
the IRAC images with TENIS-ZD1 but without all the $J$-band detected neighbors,
and the IRAC residual images.
It demonstrates that TENIS-ZD1 is well-deconvolved in the IRAC images
and the photometric effect of its bright neighbor is minimized.
The photometric redshift of TENIS-ZD1 estimated by EAZY
is $7.822^{+1.095}_{-0.725}$, with a reduced $\chi^2$=1.15.
The best-fitted templates and
the $\chi^2$ vs. redshift plots are shown in Figure~\ref{sed}.
We also fit an AGN/quasar template \citep[][]{polletta2007}
and find a best-fit redshift of $7.069^{+0.398}_{-0.411}$.
However, its $\chi^2$ is 6.50, which is higher than that of the galaxy fit.
Although an AGN is not completely ruled out, the fit favors a galaxy.

\begin{deluxetable*}{cccccccccccc}
\tabletypesize{\tiny}
\tablecolumns{16}
\tablewidth{0pt}
\tablecaption{Magnitudes of TENIS-ZD1\label{fluxes}}
\tablehead{
\colhead{U} & \colhead{B} & \colhead{R} &
\colhead{F606W} & \colhead{F850LP} &
\colhead{J} & \colhead{Ks} & \colhead{3.6$\mu{m}$} & \colhead{4.5$\mu{m}$} &
\colhead{5.8$\mu{m}$} & \colhead{8.0$\mu{m}$} }
\startdata
$>28.28$&$>28.85$&$>28.31$& 
$>28.18$&$>27.05$&
$25.12\pm.23$&$24.94\pm.34$&
$24.29\pm.10$&$23.79\pm.06$&$23.28\pm.22$&$23.58\pm.40$ \\
\enddata
\end{deluxetable*}

\section{Contamination}\label{contamination}
In this section we discuss the contaminations in our analysis.
\citet{ouchi2009} (Ouchi09, hereafter) lists all the possible sources 
of contamination for $z$-dropout studies.
Here we follow the same items as in Ouchi09 to check the
contaminations in our study.

1) Spurious sources:
One of the selection criterion of our $z>7$ sample is
that the objects must have more than $3\sigma$ detections in $J$,
and TENIS-ZD1 has a $4.3\sigma$ detection in $J$.
Statistically, an object with a $4.3\sigma$ detection is very unlikely
a spurious source.
As an additional check, we stack the images taken in 07B and 08B separately 
and generate two mosaic images for the two different semesters.
We then measure fluxes of objects in each image using the SExtractor
in the double-image mode, 
by using the $J$-band master image (07B+08B) for the source detection.
The magnitude of TENIS-ZD1 in 07B is $25.03\pm{0.29}$,
while that in 08B is $25.13\pm{0.24}$.
According to the fluxes of TENIS-ZD1 in both semesters,
it was well-detected in the two epochs.
Furthermore, TENIS-ZD1 has counterparts in $K_s$ and IRAC images.
All the above evidences suggest that TENIS-ZD1 is not a spurious source.

2) Transients:
We repeat what we did for checking spurious sources.
As we mentioned above, the flux of TENIS-ZD1 in 07B is $25.03\pm{0.29}$,
and that in 08B is $25.13\pm{0.24}$.
The magnitude change between the two epochs is only $\sim0.1$mag,
corresponding to a $\sim10\%$ flux variation.
Furthermore, TENIS-ZD1 did not move between the TENIS $J$-band (07B+08B),
$K_s$-band (09B+10B) and IRAC images,
which shows that TENIS-ZD1 is not a slow-moving solar system object.
Therefore we conclude that TENIS-ZD1 is not a transient.

3) Low-$z$ galaxies:
Our selection criteria avoid selecting galaxies at $z<6$.
However, colors of some low-$z$ galaxies could also meet the selection criteria
due to photometric errors.
We therefore discuss the possibility of that TENIS-ZD1 is a low-$z$ galaxy.
The red data point in Figure~\ref{cc-diag} shows the position of TENIS-ZD1
in the $J - 3.6\mu{m}$ vs. $F850LP - J$ diagram.
According to Figure~\ref{cc-diag},
although TENIS-ZD1 is right on the boundary of our selection criteria,
it is actually a lower-limit of its $F850LP - J$ color 
since it has no detection in the $F850LP$ image.
TENIS-ZD1 is still far away from the area of low-$z$ galaxies
even if we take its error bars into account.
Furthermore, although there is a local minimum at $z<2$ 
in the photometric redshift $\chi^2$ plot of Figure~\ref{sed},
its $\chi^2$ value is $\sim12$, 
which is much larger than that at $z\sim7-9$.
However, the color of a dusty type-2 AGN at $z\sim1.5$ can be very similar to
that of a galaxy at $z\sim7$, 
which makes it another possible source of low-$z$ contaminants.
We derived the expected fluxes at 1.4GHz, $24\mu{m}$, 
and 2-8keV for TENIS-ZD1 by assuming a type-2 AGN at $z\sim1.5$;
they are $\sim1785$uJy (radio-loud case) or $\sim0.7$uJy (radio-quiet case),
$\sim22$ mag (AB), 
and $6.4\times10^{-16} - 3.6\times10^{-15}$ ergs cm$^{-2}$ s$^{-1}$
for 1.4GHz, $24\mu{m}$, and 2-8keV, respectively.
The $1\sigma$ detection limits for the available data at these wavelengths,
however, are 8.5uJy, 21.2 mag (AB), 
and $6.7\times10^{-16}$ ergs cm$^{-2}$ s$^{-1}$ 
for 1.4GHz, $24\mu{m}$, and 2-8keV, respectively.
Non-detections of TENIS-ZD1 at these wavelengths therefore cannot rule out
a dusty radio-quiet type-2 AGN with weak X-ray luminosity at $z\sim1.5$.
Hence we did a further check by doing an SED fitting 
within a redshift range between 1.0 and 2.0 for TENIS-ZD1
using an AGN/quasar template from \citet{polletta2007}.
The best-fitted template is the QSO2 template.
According to the fitting result, 
the expected $z$-band magnitude is $\sim26$ mag,
which should be a $5\sigma$ detection in the GEMS $F850LP$ image,
but the signal-to-noise ratio of TENIS-ZD1 in the GEMS image is less than
$1\sigma$.
The best-fitted SED also fails to reproduce the steep color slope
between $K_s$ band and $4.5\mu{m}$.
Furthermore, Figure~13 in \citet{capak2011} shows 
that objects at $z\ll7$ still can match the $z>7$ selection criteria.
The $J-4.5\mu{m}$ color of TENIS-ZD1 is $1.33\pm0.24$,
which is outside the color range of any $z\ll7$ galaxies 
($J-4.5\mu{m}<1.0$ even for extreme cases).
It is therefore extremely difficult to explain TENIS-ZD1's photometric
properties using low-$z$ galaxies.

4) Galactic cool stars:
Although our selection criteria can separate dwarf stars and high-$z$
galaxies (see Figure~\ref{cc-diag-irac}),
contaminations from Galactic cool stars can still be present because of 
photometric errors.
The red data point in Figure~\ref{cc-diag-irac} indicates TENIS-ZD1.
According to Figure~\ref{cc-diag-irac},
the $3.6\mu{m} - 5.8\mu{m}$ color of TENIS-ZD1 is too red
as compared to that of dwarfs.
Moreover, the FWHM of TENIS-ZD1 in the $J$-band image shows
it is an extended source.
All these evidences suggest that TENIS-ZD1 is not a Galactic cool star.

\section{Discussions}\label{discussion}

\subsection{Surface density}
The cumulative surface density of our sample is $\sim1\times10^{-3}$
per arcmin$^2$ to $J<25.1$.
According to Figure~4 in \citet{yan2011a} (Yan11, hereafter), 
this value matches to that estimated 
from the LF based on the WFC3 $z_{850}$-dropout results for $z\sim7$ galaxies.
Yan11 claims that the estimation of the cumulative surface density 
is applicable over the redshift range of $6.4<z<7.7$.
In other words, it suggests that such an extremely luminous galaxy 
with $6.4<z<7.7$ can be found in a $\sim1000$ arcmin$^2$ survey.
Therefore, the discovery of TENIS-ZD1 is statistically predictable
based on previous studies using fainter samples.

\subsection{UV luminosity and star formation rate}
EAZY code cannot provide physical properties (e.g., age, stellar mass, etc.)
other than photometric redshifts.
Therefore we use the New-Hyperz (Rose et al.
\footnote{\url{http://www.ast.obs-mip.fr/users/roser/hyperz/}}) 
with the GALAXEV templates
\citep[][]{bc2003} to estimate the physical properties of TENIS-ZD1.
We pass the best-fitted $z_{phot}$ derived by EAZY to New-Hyperz and
treat it as a fixed parameter during the template fitting.
The extinction model from \citet{calzetti2000} is used.
The best-fitted template ($\chi^2=1.38$) 
shows that TENIS-ZD1 is a starburst galaxy, with an age of only 45M years 
and a stellar mass of $3.2\times10^{10}$ M$_\odot$.
The extinction is mild, with an $Av=0.6$.
The absolute UV magnitude M$_{1600}$ is -22.35 before the extinction correction.
We then derive the extinction-corrected M$_{1600}$, M$^{corr}_{1600}$,
using the following formula:
\begin{eqnarray}
F_{obs} = F_o \times 10^{(-0.4\times Av \times K_\lambda/4.05)}
\end{eqnarray}
\citep{calzetti2000},
where $K_\lambda$ = $K_{1600}$ = 9.97, and $Av = 0.6$.
The correction value is -1.48 mag and hence M$^{corr}_{1600}$ is -23.83.
We then estimate the SFR using the following formula \citep[][]{mpd1998}:
\begin{eqnarray}
SFR = L_{UV} (erg~s^{-1}~Hz^{-1}) / (8\times10^{27}).
\end{eqnarray}
For the extinction-corrected case, 
the estimated SFR is $\sim200$ M$_\odot$ year$^{-1}$,
while for the extinction-uncorrected case, 
the SFR is $\sim50$ M$_\odot$ year$^{-1}$.

We try to estimate the number of galaxies at z$>$7 
using the semi-analytic simulation result
from \citet{guo2011}, 
which is based on the Millennium Simulations.
The result suggests that ONE galaxy at 7$<$z$<$8
with a stellar mass greater than $10^{10}$ M$_\odot$
can be detected in the TENIS project,
which supports the existence of TENIS-ZD1.
However, the value of $\sigma_8$ used in the Millennium Simulations is 0.9,
and the most recent estimate of $\sigma_8$ is $0.809\pm0.024$
derived using the $WMAP7$ data \citep[][]{komatsu2011},
which is almost $4\sigma$ smaller than that used in the Millennium Simulations.
Structure formation would be faster with a larger $\sigma_8$
as massive objects can form earlier.
It is possible that,
with the revised $\sigma_8$, the number of galaxies similar to TENIS-ZD1
produced in the Millennium Simulation
might be less than one in our survey volume.

There are several massive galaxies at $z\sim6$ with IRAC detections reported 
in \citet{yan2005,yan2006} and \citet{eyles2007}.
According to their analyses,
the stellar masses of these galaxies are about several $10^{10}$ M$_\odot$
and their progenitors can be observed at $z\gtrsim7$.
In particular, they point out that the $z\sim6$ sample strongly indicates that
the universe was already forming galaxies as massive as $\sim10^{10}$ M$_\odot$
at $z\gtrsim7$ and possibly even at $z\sim20$,
which supports the existence of TENIS-ZD1.

Although it is challenging to explain how to accumulate so much materials
and trigger such violent star formation at $z>7$ 
with current cosmological models,
the observational results of $z\sim6$ galaxies suggest 
the existence of TENIS-ZD1 is reasonable.
Therefore bright high-$z$ objects like TENIS-ZD1 
can provide an important constraint for developing and modifying models.

\subsection{Luminosity function}
Based on the value of M$_{1600}$, 
TENIS-ZD1 is an extremely luminous object at $z>7$.
To date, TENIS-ZD1 is brighter than all published $z>7$ samples
except for the sample provided by \citet{capak2011}(Capak11, hereafter).
Hence TENIS-ZD1 provides a new constraint to the very
bright end of the LF at $z>7$.
In order to make an LF plot with TENIS-ZD1, 
we estimate the effective comoving volume for our analyses
using a Monte-Carlo simulation which is modified from Ouchi09.

First of all, we generate a mock catalogue for Lyman-Break Galaxies (LBGs)
using the starburst template of the GALAXEV model \citep[][]{bc2003}
with apparent $J$ magnitudes from 22.0 to 27.0.
The surface density vs. apparent magnitude relation of the mock catalogue
follows that of all the objects shown in Figure~\ref{completeness},
where we have assumed that 
LBGs have a similar surface density vs. apparent magnitude dependency.
The objects are uniformly distributed in the redshift space, 
and their ages are randomly distributed from 1M years to 500M years.
The extinctions of \citet{calzetti2000} with $Av=0.0$ to 2.0 are also applied
to simulate bluer and redder LBGs.
\citet{dow2007} show that roughly only one-third of 
dropout-selected galaxies at $z=5-6$ have strong Ly$\alpha$ lines 
with rest-frame equivalent widths $> 20$\AA. 
The fraction might be even lower at $z>7$ 
because of the absorption by IGM.
We therefore added Ly$\alpha$ emissions
with equivalent widths of 20\AA~ for 25\% of the objects.
We assume that 50\% of the objects have no Ly$\alpha$ emission.
For the remaining 25\% of the objects, 
we added Ly$\alpha$ emissions with equivalent widths of 1-20\AA.
After taking all of the abovementioned effects into account,
we disturb the photometry of the catalogue to match to the real data.

\begin{figure}
\epsscale{1.0}
\plotone{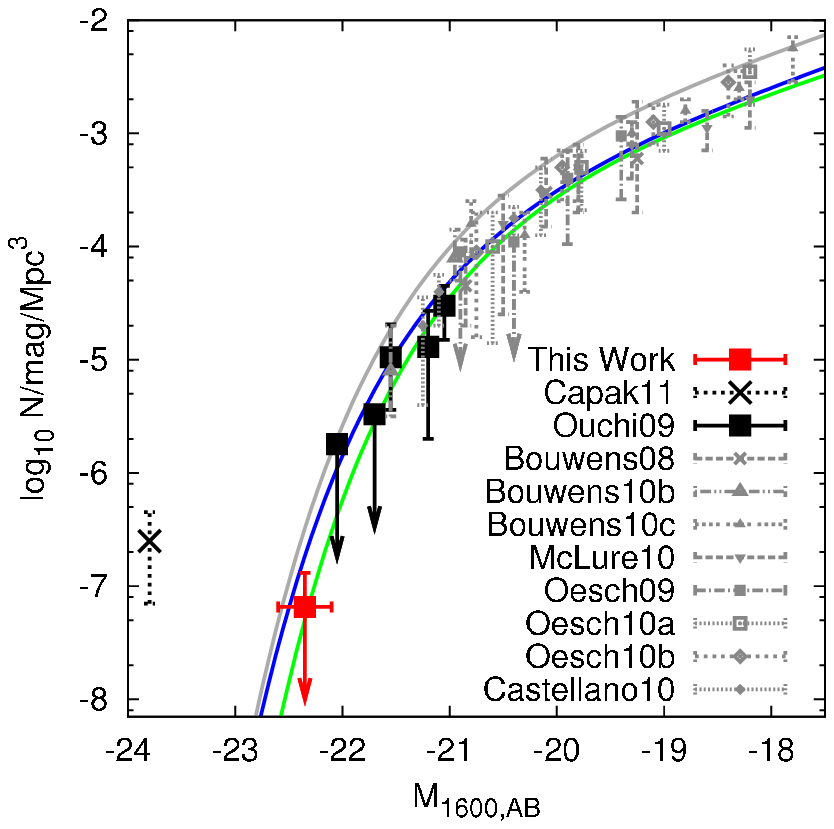}
\caption{Luminosity function at $z\sim7$.
The red filled square indicates our sample,
the black filled square indicates the sample in Ouchi09,
and the black cross indicates the sample in Capak11.
The other gray data points show the samples 
from \citet{bouwens2008,bouwens2010b,bouwens2010c,mclure2010,oesch2009,
oesch2010a,oesch2010b,castellano2010}.
The blue solid line and green solid line represent the LF 
at $z\sim7$ provided by Yan11 and Ouchi09 respectively.
The LF at $z\sim6$ in \citet{bouwens2007} is also shown
as the gray solid line.
Our result is consisitent with both Ouchi09 and Yan11.
\label{LF} }
\end{figure}

Equations~\ref{criteria1} and \ref{criteria2} are applied to
the mock catalogue to select candidates.
We computed the redshift distribution $C(m,z)$ as the ratio between
the number of candidates and the input objects in a magnitude bin
of $m=25^{+0.25}_{-0.25}$,
and estimated the comoving volume density for a given magnitude bin by
using the relation
\begin{eqnarray}
n(m)=\frac{N_{cand}(m)\times{F_c(m)}}{\int^{\infty}_0\frac{dV}{dz}C(m,z)dz},
\label{cv}
\end{eqnarray}
which is similar to Equation~2 in Ouchi09,
where $N_{cand}(m)$ is the number of candidates in the given magnitude bin,
$F_c(m)$ is the completeness correction,
$\frac{dV}{dz}$ is the differential comoving volume for the field size of the ECDFS,
and $m=25$ is the only magnitude bin we used 
since our candidate is in that magnitude bin.
In order to convert the unit of 
$n(m)$ $(N~0.5mag^{-1}~Mpc^{-3})$ to $N~mag^{-1}~Mpc^{-3}$,
a factor of 2 needs to be applied to $n(m)$.
We make the LF plot,
including the data from several previous studies, 
as shown in Figure~\ref{LF}.
The LFs provided by Ouchi09 and Yan11
fitted including luminous samples are also plotted.
The error of our data point is calculated based on a simple Poissonian
uncertainty for an event rate of 1, which is $\pm1$. 
On the other hand, the 90\% Baysian confidence level for such an event rate 
\citep[][]{kbn1991} would be 0.08-3.93. 
In either case, our result is consistent with both Ouchi09 and Yan11.

Several papers claim that the LF at $z\sim7$ decreases from $z\sim6$
at the bright end \citep[e.g.,][]{mannucci2007,castellano2010,
ouchi2009,bouwens2010b},
while Yan11 shows no/very mild evolution.
These studies are done using samples with $M_{UV}>-22$
while our sample would provide a better constraint at the very bright-end ($M_{UV}<-22$).
However, our sample is not sufficient to distinguish between the two cases
since our result is consistent with both Ouchi09 and Yan11 within the uncertainties.
On the other hand, Capak11 shows a $z\sim7$ LF with a very high bright end,
according to a sample even brighter than ours.
We note that we use very restricted criteria to select our candidate.
While these criteria would eliminate contaminations as much as possible,
some objects at $z>7$ could also be rejected at the same time.
From this point of view, the constraint that our sample provides for
the LF is a lower limit, 
and thus we cannot completely rule out the LF from Capak11.
Furthermore, cosmic variance can be as high as a factor of 10
since the extremely luminous galaxies are strongly clustered at $z>5$
\citep[][]{ouchi2004,hildebrandt2009}.
It is reasonable that the results based on observations
covering less than a few square degrees may show large discrepancies.

\section{Conclusions}\label{conclusion}
The TENIS project provides deep $J$ and $K_s$ images for the ECDFS.
We use these data with the other deep surveys in optical and infrared
to find $z>7$ objects.
New color criteria with IRAC data for selecting $z>7$ candidates are used
because we find that accurate IRAC photometry provide the key 
to resolve contamination by low-$z$ galaxies and Galactic dwarf stars.
Because the PSFs of IRAC data are much larger than those of the data
in the other wavelengths,
we introduce a novel deconvolution method to address the confusion
in the IRAC images and provide accurate IRAC fluxes.
After carefully checking our sample, 
we found one candidate at $z>7$, TENIS-ZD1.

The weighted $z_{phot}$ of TENIS-ZD1 is 7.8,
with an estimated stellar mass = $3.2\times10^{10}$ M$_\odot$.
The extinction-corrected current SFR of TENIS-ZD1 is 
$\sim200$ M$_\odot$ year$^{-1}$,
but must have been very much higher in the past.
We summarize the results of our sample below:
1) The discovery of TENIS-ZD1 is predictable in an 1,000 arcmin$^2$ survey
according to the surface density derived from literatures.
2) Our sample matches to both LFs from Ouchi09 and Yan11.
3) While the existence of TENIS-ZD1 is supported 
by the observations at $z\sim6$,
it is still hard to explain 
how such a massive galaxy can form in the early universe
based on current cosmological models for structure formation.

In the end, we would like to emphasize that
the TENIS project provides the most comprehensive dataset possible 
to address the problem of finding luminous high-$z$ galaxies.
Compared with fainter high-$z$ samples,
such high-$z$ massive objects like TENIS-ZD1 provide the greatest leverage
in testing galaxy formation/evolution models.
Such objects are the best candidates for follow-up spectroscopy
with large telescopes;
at the same time, to minimize wastage of observing times on large telescopes,
all efforts must first be made to weed out contaminants.
The sizes of the $z\sim7$ bright samples are very small 
due to the limited field size of deep near-infrared surveys,
and the purity and completeness issues of the dropout technique 
need to be further investigated.
Due to the severe contamination issue of the bright end of the $z\sim7$ LF,
all the studies make great efforts on increasing the purity of their samples,
however, all these efforts may also harm the completeness of the samples,
which makes the bright end of the $z\sim7$ LF seriously under-estimated.
\citet{hc2006} point out that the bright end of the LF at $z\sim6$
based on LAE studies is much higher than that based on LBG studies.
Capak11 also claims that the bright end of the LF at $z\sim7$
could be much higher than the results derived using the dropout technique.
The bright samples will increase very soon 
since there will be more and more large field near-infrared surveys.
However, simply increasing the bright sample size cannot solve the bias issue.
The best way is to confirm their redshifts spectroscopically.
Each spectroscopically confirmed object will set a robust lower-limit
for that luminosity bin of the LF.
The James Webb Space Telescope (JWST) and 30m-class telescopes (e.g., TMT,
GMT) will be certainly capable for these confirmation observations.
But before the era of JWST and 30m-class telescopes comes,
the current 10m-class telescopes are still able to perform the
observations for objects with $M_{UV}<-21$ 
as long as strong Ly$\alpha$ lines exist.
These spectroscopic observations will provide very important information
for inspecting the real performance of the dropout studies at $z>7$.

\acknowledgments
We thank the referee for comments that greatly improve the manuscript.
We are grateful to the CFHT staff for help on obtaining the data,
and H. K. C. Yee and Y. T. Lin for useful discussions.
This paper is based on observations obtained with WIRCam, 
a joint project of CFHT, Taiwan, Korea, Canada, France, 
and the Canada-France-Hawaii Telescope (CFHT) 
which is operated by the National Research Council (NRC) of Canada, 
the Institute National des Sciences de l'Univers of 
the Centre National de la Recherche Scientifique of France, 
and the University of Hawaii.
Access to the CFHT was made possible by the Ministry of Education,
the National Science Council of Taiwan as part of the Cosmology and 
Particle Astrophysics (CosPA),
the Institute of Astronomy and Astrophysics, Academia Sinica, 
and National Tsing Hua University, Taiwan.
We gratefully acknowledge support from the National Science Council of Taiwan
grant 98-2112-M-001-003-MY2 (W.H.W.), 99-2112-M-001-012-MY3 (W.H.W),
and Japanese Government FIRST program (H.K.).

\end{document}